\input harvmac
\input scrload


\input labeldefs.tmp

\writedefs

\def\tpap{(2 \pi \sqrt{\alpha'})}


\Title{ \vbox{\baselineskip12pt \rightline{HUTP-01/A047}
 \rightline{YITP-SB-01-55}
 \rightline{hep-th/0110109} 
 \vskip -.25in} }{ \vbox{
\centerline{Inside an Enhan\c con:} 
\centerline{Monopoles and Dual Yang-Mills Theory} } }

\centerline{ Martijn Wijnholt$^1$ and Slava Zhukov$^2$}
\vskip .1in
\centerline{$^1$ \sl Jefferson Laboratory of Physics, Harvard
University}
\centerline{\sl Cambridge, MA 02138, USA}
\centerline{\it wijnholt@fas.harvard.edu }

\vskip .1in

\centerline{$^2$ \sl C.N. Yang Institute for Theoretical Physics}
\centerline{\sl SUNY, Stony Brook, NY 11794-3840, USA}
\centerline{\it zhukov@insti.physics.sunysb.edu}

\vskip 0.3in

\centerline{\bf Abstract}

\noindent
In this paper we study wrapped brane configurations that give rise to
three dimensional pure Yang-Mills theory with eight supercharges. The
corresponding supergravity solution is singular and it was conjectured
that the singularity is removed by an enhan\c con
mecha\-nism. Instead, we incorporate non-perturbative gauge fields
into supergravity and find a smooth solution for these
configurations. Along the way we derive a non-abelian super\-gravity
Lagrangian for type IIA on K3 and explicit formulae for the toroidal
reduction of the heterotic string with non-abelian gauge fields. We
proceed to analyse the duality with Yang-Mills theory and find that
the dual background is a monopole configuration in little string
theory.

\Date{October 2001}


\lref\MaharanaMY{
J.~Maharana and J.~H.~Schwarz,
``Noncompact symmetries in string theory,''
Nucl.\ Phys.\ B {\bf 390}, 3 (1993)
[hep-th/9207016].
}

\lref\HL{
J.~A.~Harvey and J.~Liu,
``Magnetic monopoles in N=4 supersymmetric low-energy superstring theory,''
Phys.\ Lett.\ B {\bf 268}, 40 (1991).
}

\lref\JPP{
C.~V.~Johnson, A.~W.~Peet and J.~Polchinski, ``Gauge theory and the
excision of repulson singularities,'' Phys.\ Rev.\ D {\bf 61}, 086001
(2000) [hep-th/9911161].
}

\lref\Pol{
J.~Polchinski, ``String Theory,'' Vol. 1 \& 2, Cambridge University
Press, 1998
}

\lref\Sen{
A.~Sen, ``Strong - weak coupling duality in four-dimensional string
theory,'' Int.\ J.\ Mod.\ Phys.\ A {\bf 9}, 3707 (1994)
[hep-th/9402002].
}

\lref\SenCJ{
A.~Sen, ``String string duality conjecture in six-dimensions and
charged solitonic strings,'' Nucl.\ Phys.\ B {\bf 450}, 103 (1995)
[hep-th/9504027].
}

\lref\WittenEX{
E.~Witten, ``String theory dynamics in various dimensions,'' Nucl.\
Phys.\ B {\bf 443}, 85 (1995) [hep-th/9503124].
}

\lref\KiritsisZI{
E.~Kiritsis, N.~A.~Obers and B.~Pioline, ``Heterotic/type II triality
and instantons on K3,'' JHEP {\bf 0001}, 029 (2000) [hep-th/0001083].
}

\lref\Prasad{
M.~K.~Prasad, ``Exact Yang-Mills Higgs Monopole Solutions Of Arbitrary
Topological Charge,'' Commun.\ Math.\ Phys.\ {\bf 80}, 137 (1981).
}

\lref\PR{
M.~K.~Prasad and P.~Rossi,
``Construction Of Exact Multi - Monopole Solutions,''
Phys.\ Rev.\ D {\bf 24}, 2182 (1981).
}

\lref\VafaFJ{
C.~Vafa and E.~Witten, ``A One loop test of string duality,'' Nucl.\
Phys.\ B {\bf 447}, 261 (1995) [hep-th/9505053].
}

\lref\StromingerCZ{
A.~Strominger, ``Massless black holes and conifolds in string
theory,'' Nucl.\ Phys.\ B {\bf 451}, 96 (1995) [hep-th/9504090].
}

\lref\str{
A.~Strominger,
``Heterotic Solitons,''
Nucl.\ Phys.\ B {\bf 343}, 167 (1990)
[Erratum-ibid.\ B {\bf 353}, 565 (1990)].
}

\lref\VafaZH{
C.~Vafa,
``Gas of D-Branes and Hagedorn Density of BPS States,''
Nucl.\ Phys.\ B {\bf 463}, 415 (1996)
[arXiv:hep-th/9511088].
}

\lref\BershadskyQY{
M.~Bershadsky, C.~Vafa and V.~Sadov,
``D-Branes and Topological Field Theories,''
Nucl.\ Phys.\ B {\bf 463}, 420 (1996)
[hep-th/9511222].
}

\lref\VafaBM{
C.~Vafa,
``Instantons on D-branes,''
Nucl.\ Phys.\ B {\bf 463}, 435 (1996)
[arXiv:hep-th/9512078].
}

\lref\KalloshYZ{
R.~Kallosh and A.~D.~Linde,
``Exact supersymmetric massive and massless white holes,''
Phys.\ Rev.\ D {\bf 52}, 7137 (1995)
[hep-th/9507022].
}

\lref\MaldacenaMW{
J.~Maldacena and C.~Nunez,
``Supergravity description of field theories on curved manifolds and a no  go theorem,''
Int.\ J.\ Mod.\ Phys.\ A {\bf 16}, 822 (2001)
[hep-th/0007018].
}

\lref\MaldacenaYY{
J.~M.~Maldacena and C.~Nunez,
``Towards the large N limit of pure N = 1 super Yang Mills,''
Phys.\ Rev.\ Lett.\  {\bf 86}, 588 (2001)
[arXiv:hep-th/0008001].
}

\lref\OoguriWJ{
H.~Ooguri and C.~Vafa,
``Two-Dimensional Black Hole and Singularities of CY Manifolds,''
Nucl.\ Phys.\ B {\bf 463}, 55 (1996)
[hep-th/9511164].
\hfill \break
D.~Kutasov,
``Orbifolds and Solitons,''
Phys.\ Lett.\ B {\bf 383}, 48 (1996)
[hep-th/9512145].
}

\lref\HananyIE{
A.~Hanany and E.~Witten, ``Type IIB superstrings, BPS monopoles, and
three-dimensional gauge dynamics,'' Nucl.\ Phys.\ B {\bf 492}, 152
(1997) [hep-th/9611230].
}

\lref\SeibergZK{
N.~Seiberg, ``New theories in six dimensions and matrix description of
M-theory on T**5 and T**5/Z(2),'' Phys.\ Lett.\ B {\bf 408}, 98 (1997)
[hep-th/9705221].
}

\lref\AharonyTI{ For a review, see \hfill \break
O.~Aharony, S.~S.~Gubser, J.~Maldacena, H.~Ooguri and Y.~Oz,
``Large N field theories, string theory and gravity,''
Phys.\ Rept.\  {\bf 323}, 183 (2000)
[hep-th/9905111], and references therein.
}

\lref\GibbonsZT{
G.~W.~Gibbons and P.~K.~Townsend,
``Antigravitating BPS monopoles and dyons,''
Phys.\ Lett.\ B {\bf 356}, 472 (1995)
[arXiv:hep-th/9506131].
}

\lref\JohnsonBM{
C.~V.~Johnson,
``Enhancons, fuzzy spheres and multi-monopoles,''
Phys.\ Rev.\ D {\bf 63}, 065004 (2001)
[arXiv:hep-th/0004068].
}

\lref\WZ{
M.~P.~Wijnholt, S.~Zhukov, in preparation.
}

\lref\WeinbergEQ{
E.~J.~Weinberg and A.~H.~Guth, ``Nonexistence Of Spherically Symmetric
Monopoles With Multiple Magnetic Charge,'' Phys.\ Rev.\ D {\bf 14}, 1660
(1976).
}

\lref\KlebanovHB{
I.~R.~Klebanov and M.~J.~Strassler,
``Supergravity and a confining gauge theory: Duality cascades and  chiSB-resolution of naked singularities,''
JHEP {\bf 0008}, 052 (2000)
[arXiv:hep-th/0007191].
}

\lref\GiveonPX{
A.~Giveon and D.~Kutasov,
``Little string theory in a double scaling limit,''
JHEP {\bf 9910}, 034 (1999)
[arXiv:hep-th/9909110].
}

\lref\BachasUM{
C.~P.~Bachas, P.~Bain and M.~B.~Green,
``Curvature terms in D-brane actions and their M-theory origin,''
JHEP {\bf 9905}, 011 (1999)
[arXiv:hep-th/9903210].
}

\newsec{Introduction}

\subsec{Overview}

The basic ingredient of gauge/gravity dualities is the existence of two
pictures representing the same physical system, a supergravity picture and
a D-brane picture. After the original dualities found by Maldacena and
others \AharonyTI, recently a lot of attention has turned to extending the
list of examples to settings with less than maximal supersymmetry. In one
such attempt, Johnson, Peet and Polchinski \JPP\ considered a brane setup
that could potentially lead to a useful gravity dual for pure ${\scr N} = 4$
Yang-Mills theory in three dimensions and related cases with eight
supercharges. Curiously the candidate gravity solution representing these
branes had a naked singularity. This problem was addressed in \JPP\ by showing
that an unexpected phenomenon had not been taken into account: at a special
radius, called the enhan\c{c}on radius, there are extra massless fields
that enhance the abelian gauge symmetry of the naive solution to
$SU(2)$. It was then proposed that the brane constituents could not be
moved below the enhan\c con radius, but instead should be spread on a
spherical shell at this radius. Inside the shell, the geometry with the
singularity is excised and replaced by flat space.

In this article we  directly incorporate the effect of gauge
symmetry enhancement.  We introduce light non-abelian fields in the
supergravity Lagrangian, which we derive in appendix B, and ask for a BPS
solution to the equations of motion of this Lagrangian. Such a solution
exists and has the correct asymptotic behaviour to represent the system of
wrapped D-branes. This new gravity solution is automatically non-singular,
and differs from the shell of branes that was argued to be the correct
supergravity solution in \JPP.

We explain in section two how one may find this new solution. It turns
out there is a different solution, the Harvey-Liu monopole \refs{\HL,
\GibbonsZT}, that can be mapped to the desired one via non-abelian
toroidal reduction and S-duality.  In order to perform the reduction,
we have worked out explicit formulae in appendix A which generalise
those of Maharana and Schwarz \MaharanaMY.  There is a surprise along
the way, since the Harvey-Liu monopole has an excited $H$-field, yet
the solution we seek is not an $H$-monopole.

In section three, we analyse the limits that are involved in
decoupling the Yang-Mills theory living on the D-branes. We then
perform the corresponding limits on the supergravity solution, and
state the duality we arrive at. The gravitational modes turn out to be
a red herring; they can be decoupled without affecting the Yang-Mills
modes.  Finally in section four we summarise and add some more
relevant remarks.

For the decoupling limit we consider a configuration of wrapped branes
that is different from \JPP. Although they are related by T-duality,
there appears to be a discrepancy between the values of the three
dimensional Yang-Mills coupling in the two cases. However, they can be
reconciled by adding a correction to the D-brane action. We discuss
this correction in section four also.

\subsec{Review}

One particular setup considered in \JPP\ is that of $N$ D6 branes in type
IIA string theory. To reduce the number of supercharges from sixteen to
eight, these D6 branes are wrapped on a K3 surface. In an appropriate low
energy limit one can decouple the low energy Yang-Mills theory living on
the D6 branes. If one also decouples the Kaluza-Klein modes on the K3 one
is left over with a three dimensional ${\scr N} = 4$ gauge theory along the
non-compact directions of the branes. An ${\scr N} = 4$ vector multiplet has
three real scalars, which are accounted for by reduction of the three
scalars in the seven dimensional Yang-Mills multiplet living on a
D6-brane. In the zero instanton sector the gauge field has no zero modes on
the K3 and therefore there are no additional scalars, hence no
hypermultiplets. So the theory of interest is three dimensional ${\scr N} =
4$ pure Yang-Mills theory. One can get additional matter multiplets by
considering non-zero instanton number in the D6-brane gauge theory, which
is the same as adding extra D2 branes, or by adding D4 branes wrapped on a
curve inside the K3 with genus greater than one. As a further
generalisation, other pure gauge theories with eight supercharges, such as
${\scr N} = 2$ Yang-Mills theory in 3+1 dimensions, may be obtained by
wrapping branes of different dimensionalities on the K3 surface. However we
restrict to the case of three dimensional pure Yang-Mills theory in this
paper.

The compactification on the K3 surface yields an interesting effect
\BershadskyQY : because of a tr($R \wedge R$) term in the Chern-Simons
action of the D-branes, which is non-zero when the K3 is part of the
worldvolume, the D6 brane acts as a source for the three form Ramond
field. Based on the relation with the heterotic string on $T^4$ it was
conjectured that we should associate an anti D2 brane charge to it,
rather than a D2 brane charge. We will refer to the wrapped D6 as
D6-$\overline{\rm D2}$. Similarly a D4 brane will also carry an
induced charge when wrapped on the K3 surface, and we write it as
D4-$\overline{\rm D0}$.

There is another known $R^2$ term in the D-brane action, written down
explicitly in \BachasUM. Its physical effect is a correction to the
tension of any brane which is wrapped on a K3. In fact for, say, the D6
brane considered above, this correction reduces its tension exactly by
the tension of a D2 brane, as one would have expected for a BPS
object with the charges of D6-$\overline{\rm D2}$.

A supergravity solution for D6-$\overline{\rm D2}$ was written down in
\JPP\ using the harmonic function rule. The solution exhibits an attractor
flow as a function of the radius: as one comes in from infinity, the
volume of the cycle on which the branes are wrapped, in this case the
K3 itself, shrinks. The volume first reaches the ``self-dual'' value
$\tpap^4$, and after that the metric coefficient $g_{tt}$ blows up at
finite radius as the volume shrinks to zero size.  This singularity is
referred to as the repulson, for the repulsive behaviour it exhibits
at small radii \KalloshYZ.  It defies an explanation in terms of
D-brane physics, for instance it does not satisfy the criterion of
\MaldacenaMW\ for it to be interpreted in terms of a dual field
theory, and the force between a massive uncharged particle and a D6
brane ought to be attractive.  The presence of this singularity
therefore casts doubt on whether the repulson geometry is the correct
supergravity representation of a collection of wrapped D6 branes.

An important observation in \JPP\ was that the D4 brane becomes massless
when wrapped on a K3 surface of self-dual volume due to the cancelling
tensions of the D4 brane and the anti D0 brane. In fact the wrapped D4
brane is a W-boson and, as can be seen by duality with the heterotic string
on $T^4$, its masslessness signals that the abelian gauge symmetry under
which the D6 is charged is enhanced to SU(2). The value of the radius
where the K3 is at self-dual volume is then fittingly called the enhan\c
con radius.

The appearance of new massless fields violates the basic premise that
the supergravity Lagrangian incorporates all the light fields. We
therefore need a more general Lagrangian, for which the equations of
motion are more likely to have a finite energy solution without a
singularity, or at least with a reasonable singularity. It is not
known how to write such a Lagrangian for type IIA supergravity by
compactifying on a K3 surface. We in fact work it out by S-duality
with the heterotic string, where the compactification is on a $T^4$
and the non-abelian fields are perturbative.

To find the correct supergravity solution, we use the observation in \JPP\
that a wrapped D6 brane must be represented by a monopole, since it is the
electric-magnetic dual of a wrapped D4 brane, which is a W-boson. Again it
would be easier to appeal to S-duality and find a supergravity monopole in
heterotic string theory. In fact, a promising candidate was written down by
Harvey and Liu \HL\ based on the gauge fivebrane Ansatz. We will write this
solution in terms of the toroidally reduced six dimensional fields and then
S-dualise it. The resulting supergravity solution is a smooth monopole and
represents the wrapped D6 branes.

The authors of \JPP\ suggested a different modification of the
geometry. They noticed that a probe brane in the repulson background
acquires a potential that prevents it from moving inside the enhan\c con
radius. Therefore they proposed a supergravity solution in which the D6
branes are still present, except that they are no longer placed at the
origin but on a spherical shell at the enhan\c con radius. While in our
geometry the energy density is not localised at the origin, we do not see
it localised on such a spherical shell either.

The relation between the enhan\c con and BPS monopoles has also been
explored in \JohnsonBM, although from a different point of view.

\newsec{A supergravity monopole}

\subsec{The Harvey-Liu monopole}

As explained in section one, we would like to start with a monopole
configuration in heterotic string theory. We will use
normalisations and conventions of \Pol. The ten dimensional string
frame action is in this case
\eqn\SHetFull{ S = {2\pi \over \tpap^8}
             \int d^{10}x \, \sqrt{-g} e^{-2\Phi}
    \Big\{ R + 4 (\partial \Phi)^2 - {1\over 12} H^2
    - {\alpha'\over 8} \tr(F^2) + \ldots \Big\} .}
The ellipses indicate that we have dropped fermionic and higher order
terms. The definition of the three-form $H$ is such that it includes
the Chern-Simons term:
\eqn\dH{ d H = -{\alpha' \over 4}\tr\big(F \wedge F\big)
\ .}
The gauge group in the heterotic theory is well known to be restricted
by the anomaly to be either $SO(32)$ or $E_8 \times E_8$. The traces
above should be taken in the vector representation of those groups. In
what follows only an $SU(2)$ subgroup will play a role and therefore
we will focus on it as our gauge group and forget the rest of the
generators of $SO(32)$ or $E_8 \times E_8$.

Following earlier work on the gauge fivebrane, Harvey and Liu
\HL\ (see also \GibbonsZT) found a supergravity monopole by starting with a
flat space BPS monopole and dressing it up with a non-trivial metric,
dilaton and $H$-field.  The similarity with the heterotic gauge fivebrane
becomes clear if one adopts the convention of writing the Bogomol'nyi
equations for BPS monopoles as Yang-Mills self-duality equations on ${\bf
R}^3 \times S^1$, so that $A_6$ plays the r\^{o}le of the Higgs field. It
was found in
\str\ that the heterotic BPS equations are solved by the following Ansatz,
written in terms of the ten dimensional fields:
\eqn\monopole{
\eqalign{ F_{pq} &= \pm {1\over 2}{\epsilon_{pq}}^{rs} F_{rs} \cr
H_{pqr} &= \mp 2 \, {\epsilon_{pqr}}^s \partial_s \Phi \cr
g_{\mu\nu} &= {\rm diag}\ (-1,  e^{2\Phi}, e^{2\Phi}, e^{2\Phi}, 1, 1,
         e^{2\Phi}, 1, 1, 1) \cr
\nabla_i \nabla^i \Phi &=
\mp {\alpha'\over 8} \,  \epsilon^{pqrs}
        \, \tr\, F_{pq} F_{rs} . } }
In these equations indices are raised with the metric $g_{\mu\nu}$.
We have introduced a number of conventions that we will
use through the rest of the paper. We summarise them below:
\medskip
\item{$\cdot$} $\mu, \nu, \lambda$ are space-time indices in ten and,
later, six dimensions;
\item{$\cdot$} $\alpha,\beta,\gamma \in \{6,7,8,9\}$ are the internal 
indices for the four-torus;
\item{$\cdot$} $a,b,c$ or $\parallel$
will be used for 0, 4 and 5, the directions
along the soliton worldvolume;
\item{$\cdot$} $i,j,k$ or $\perp$ are for the transverse directions 1, 2 and 3;
\item{$\cdot$} $p,q,r,s$ stand for 1, 2, 3 and 6, in order to
write the self-duality equation;
\item{$\cdot$} $I,J,K$ are indices for the adjoint representation of
the gauge group SU(2).
\medskip

\noindent
Apart from $A_1, A_2, A_3$ and $A_6$ all components of the SU(2) gauge
field are set to zero.
This solution preserves half of the sixteen supercharges. It is not an exact
solution of string theory; the metric has to be adjusted order by order in
$\alpha'$. To understand what charges the soliton carries we have to write down
the solution in terms of the six dimensional fields, which is done in the
next subsection.

For later use let us write down the expressions for a single monopole from
Harvey and Liu \HL. The vector $\vec{r}$ will denote $(x_1,x_2,x_3)$ and $r$ is
its length. The gauge fields are then just those of a flat space
monopole
\eqn\GaugeHiggsOne{  A^J_i(\vec{r}) = \epsilon_{iJK} 
{x^K\over r^2}(K(Cr) - 1), \quad A_6^I(\vec{r}) = {x^I\over r^2}H(Cr) }
The functions $K(x)$ and $H(x)$ are given by
\eqn\HKdefn{ H(x) \equiv x \coth (x) - 1, 
			\quad K(x) \equiv {x\over \sinh (x)} . }
Here $C = \sqrt{\left<A_6^I A_6^I \right>}$ is the expectation value of
the Higgs field at infinity.  The dilaton can be found by solving a
Laplace equation with a flat space Laplacian
\eqn\LaplaceFlat{ \eqalign{\partial_i \partial_i e^{2\Phi}
 &= - {\alpha'\over 8} F^I_{pq} F^{I}_{pq} \cr
 &=  - {\alpha' \over 2 r^4}(2 H^2 K^2 + (K^2 - 1)^2)\ ,}}
which results in
\eqn\singlemonopoledilaton{
\eqalign{e^{2\Phi}
  &=  e^{2\Phi_0} + { \alpha'\over 4 r^2}(1 - K(Cr)^2 + 2 H(Cr)) \cr
  &=     e^{2\Phi_0} + {\alpha'\over 4} (C^2 - {H(Cr)^2\over r^2})\ .}}

For higher charge monopoles explicit expressions are hard to come
by. But one can do the next best thing by assuming a solution to the
flat space Bogomol'nyi equations for monopoles. For this purpose, let
us introduce a fiducial flat space BPS monopole solution of charge $N$
labelled by ${\scr A}(\vec r)$ and $h(\vec r)$, such that $h^2(\infty)
= 1$, and try to express all the ten dimensional fields in terms of
these. Notice that $\scr A$ and $h$ depend on $N$ together with the
choice of a point on the monopole moduli space. But in order to
simplify notation we do not show this dependence explicitly. The ten
dimensional gauge and Higgs fields for the Harvey-Liu monopole can be
written as
\eqn\aa{ A_i(\vec{r}) = C{\scr A}_i(C\vec{r}), \quad A_6(\vec{r}) = C\,  
          h(C\vec{r}) .}
Next, we can get a useful expression for the dilaton by using the flat
space BPS equations $ D_i A_6^I = B^I_i = - (1/2) \epsilon_{ijk}
F^I_{jk}$, the Bianchi identity $D_i B^I_i = 0$ and the Ansatz \monopole:
\eqn\aa{ \eqalign{
 \partial_i \partial_i e^{2\Phi} &= 
	  -{\alpha'\over 8}{\rm tr} (F_{pq} F_{pq}) \cr
	&= -{\alpha'\over 2} {\rm tr}(D_i A_6\, D_i A_6) 
	= -{\alpha'\over 2}\partial_i {\rm tr}(A_6 D_i A_6) \cr
	&= -{\alpha'\over 4}\partial_i \partial_i {\rm tr}(A_6\, A_6)
 .}
}
It follows that  $e^{2\Phi}$ and $-  \alpha' A_6^2/4$ can only differ by a
constant. Since $A_6^2 \to C^2$ as $r$ goes to infinity, the relation between
the dilaton and the Higgs field in
general is
\eqn\monopoledilaton{ \eqalign{
e^{2\Phi} &= e^{2\Phi_0} + { \alpha' C^2\over 4} \left(1 - {A_6^2(\vec{r})
\over C^2}\right) \cr
   &= e^{2\Phi_0} + { \alpha' C^2\over 4} \left(1 - h^2(C\vec{r})\right) .}}
This indeed reproduces the result for a single monopole listed in
\singlemonopoledilaton.

Knowing the dilaton $\Phi$, from \monopole\ we immediately know the
metric and the $H$ field. However, for the later use we also need the
$B$ field. First, from \monopole\ we compute the only non-zero
component of $H$ :
\eqn\Hijsix{
 \eqalign{ H_{ij6} &= - \epsilon_{ijk} \partial_k e^{2\Phi} 
   = {\alpha'\over 4} \epsilon_{ijk}\partial_k {\rm tr}(A_6\, A_6) 
   = -{ \alpha'\over 2} \tr(A_6 F_{ij}) \cr
 & = - {\alpha'\over 4} \omega^{cs}_{ij6}
    - {\alpha'\over 4} \partial_i \tr(A_j A_6)
    + {\alpha'\over 4} \partial_j \tr(A_i A_6)\ .} }
Here $ \omega^{cs}_{ij6}$ is the non-abelian Chern-Simons form built from
the gauge and Higgs fields entering the definition of $H$, see
\CHdefn. Then we may take%
\foot{$B$ is not gauge independent and this equation amounts to choosing a
specific gauge.}
\eqn\Bisix{ B_{i6} = -{\alpha'\over 4} \tr(A_i A_6)\ . }
Furthermore, from \monopole\ we find that $H_{ijk} = 0$, hence
\eqn\dB{ dB_{ijk} = {\alpha'\over 4} \omega^{cs}_{ijk} .}
From this one may write an integral expression for $B_{ij}$ in terms
of the gauge fields, but we will have no need for it here.

The Higgs expectation value breaks the $SU(2)$ gauge symmetry, and so 
the mass of the W-boson in six dimensions depends
on the choice of $C$. The W-boson is a BPS state of
the heterotic string and its mass is therefore determined by the
rightmoving momentum of the string. The latter in turn can be
expressed in terms of the background fields on the torus and the
charges it carries under the remaining $U(1)$'s. The W's have charge
$\pm 1$ under the $U(1)$ that remains after $SU(2)$ is broken, and
they get a mass due to non-zero Wilson lines, so we obtain (\Pol,
Vol. II, Pg. 77)
\eqn\aa{ M^2_W =  k_R^2 = (\pm 1)^2 G^{\alpha\beta} A^I_\alpha A^I_\beta .}
Since only $A_6$ is nonzero and $A_6^2 \to C^2$ as $r$ goes to infinity,
we get
\eqn\hetWmass{ M_W = e^{-\Phi_0} C ,}
a relation that will be useful later on.

\subsec{Non-abelian toroidal reduction}

Our next step is to compactify the ten-dimensional heterotic theory on
a four-torus and rewrite the monopole solutions of the previous
subsection in terms of the six-dimensional fields. Because of the
nature of the monopole, this involves a toroidal reduction in the
presence of non-abelian gauge fields. Following Maharana and Schwarz
\MaharanaMY\ we perform such a reduction in appendix A.

There we introduce the following notation for the fields of the resulting
six-dimensional theory. From the ten-dimensional metric we obtain the six
dimensional metric $g_{\mu \nu}$, four $U(1)$ gauge fields
$A^{(1)\alpha}_\mu$ and a $4 \times 4$ symmetric matrix of scalars
$G_{\alpha \beta}$.  From the dilaton we obtain its six-dimensional cousin,
$\phi$.  The ten-dimensional two-form field gives rise to a two-form
$B_{\mu \nu}$ in six dimensions, four more $U(1)$ gauge fields
$A^{(2)}_{\mu \alpha}$ and a $4 \times 4$ antisymmetric matrix of scalars
$B_{\alpha \beta}$. Finally, the ten-dimensional gauge fields produce
the non-abelian gauge fields in six dimensions $A^{(3)I}_\mu$ and four adjoint
scalars $a^I_\alpha$.

We choose the torus directions to run from 6 to 9. Now we use formulas
from the appendix and write down the six-dimensional monopole
solution.

The metric in ten dimensions is diagonal \monopole, which makes it
easy to choose a vielbein. Then from \fullvielbein\ and \fullmetric\
we have in six dimensions
\eqn\rdcmonsol{\eqalign{
&g_{\mu \nu} = {\rm diag}
  (-1,e^{2 \Phi},e^{2 \Phi},e^{2 \Phi},1,1), \quad
G_{\alpha \beta} =  {\rm diag}
  (e^{ 2 \Phi},1,1,1), \cr
&A^{(1)\alpha}_\mu = 0 \ .}}
For the six-dimensional dilaton we have from \CDdilaton\
\eqn\CDdilaton{e^{-2 \phi} = 
{V_{T^4} \over \tpap^4} e^{-2 \Phi} =
{(\int_{T^4} dy) \,  e^\Phi \over \tpap^4} e^{-2 \Phi} =
\lambda \, e^{- \Phi}\ , }
where we have denoted $\lambda = \int_{T^4} dy / \tpap^4 $ the
``coordinate'' volume of the internal torus in $\alpha'$ units.  This
specific dilaton normalization will turn out very convenient later,
when we will apply S-duality.

The ten-dimensional gauge fields reduce to the expected monopole
configuration for the $A^{(3)I}_\mu$ while the role of the Higgs field
is played by $a^I_6$, see \CAdefn
\eqn\Athree{ A^{(3)}_i(\vec r) = C{\scr A}_i(C\vec{r}),  \quad a_6(\vec{r}) = C\,  h(C\vec{r}) .}

The reduction of the $B$ field is the most interesting. In
ten dimensions the Harvey-Liu monopole has non-vanishing $H$ field
\monopole\ and carries an $H$-monopole charge. Since $H_{ij6}$ is non-zero
one might have expected the six-dimensional solution to carry an
$H$-monopole charge also.  However we are looking for a pure non-abelian
monopole; we do not want the soliton to carry charge under a second gauge
field. 

In order to see if there is a six dimensional $H$-charge or not we
need to examine the corresponding six dimensional gauge fields,
$A^{(2)}_{\mu\alpha}$. It was already found by Sen \Sen\ for the
Harvey-Liu monopole compactified to four dimensions that there is no
charge under $A^{(2)}$ gauge fields.  In our case we can also see
explicitly from \CDBdefn\ in appendix A and \Bisix\ that all
contributions to $A^{(2)}_{\mu\alpha}$ cancel:
\eqn\AtwoFtwo{ A^{(2)} = 0 , \quad F^{(2)} \equiv 0 .}
So we conclude that the soliton in fact has no six dimensional $H$-monopole
charge.

Finally, from
\CDBdefn\ it is also clear that the six-dimensional scalars $B_{\alpha
\beta}$ vanish, while the two-form $B_{\mu \nu}$ is still determined
by \dB. As a result, the contributions due to $dB$ and the Chern-Simons
term to $H_{\mu\nu\lambda}$ also cancel after compactification, and we have
$H_{\mu\nu\lambda}=0$ in six dimensions.

\subsec{S-duality between type IIA on K3 and heterotic on $T^4$}

In this subsection we will use S-duality between heterotic string theory
compactified on $T^4$ and type IIA string theory on K3.  First we will
obtain the Lagrangian of the latter theory near the point of enhanced gauge
symmetry and then we will rewrite the monopole solutions of the previous
subsection in the type IIA language. Let us begin by explaining how we are
going to apply the duality. 

S-duality is believed to be an exact equivalence of the two theories.  Its
simplest manifestation arises at the level of the low-energy effective
actions. Usually these actions are written at the generic points on the
moduli space, far away from the loci where enhanced gauge symmetry occurs.
In particular it means that on the heterotic side the gauge group is
maximally broken by Wilson lines along the internal directions of the torus
and in addition none of the torus radii are close to the self dual
value. On the type IIA side it means that the volumes of all the homology
cycles should be large in $\alpha'$ units. At such points in the moduli
space the low energy effective action involves only the massless abelian
fields. It can be expanded in series in the number of derivatives and also
in the powers of the coupling constant - the dilaton.

At the two-derivative level the actions can be completely determined
either by a direct compactification from ten dimensions or from
supersymmetry alone. Then a simple change of variables brings the
two-derivative effective actions into one another \refs{\SenCJ,
\WittenEX}. Under this change of variables the dilaton changes sign
and the coupling constant is inverted. The string metric is rescaled
by the power of the dilaton and because of that the masses measured by
``natural'' observers in the two theories are also rescaled. The
six-dimensional two-form field $B_{\mu \nu}$ goes into its Hodge dual
up to a power of the dilaton.  Finally, all the scalar and gauge
fields are the same in both theories.

Beyond the two-derivative level, the identification of the effective
actions is much more problematic. Since the coupling constants are
inverted under S-duality, the corresponding expansions run around
different points. Therefore to match a given term on one side one may
potentially need to know all the terms on the other. Nevertheless,
certain very special terms have been computed and perfectly matched on
both sides \KiritsisZI.

Apart from the effective actions, the duality can be further probed by
comparing charges and masses of the BPS states in the two
theories. The virtue of the BPS states is that they exist at all
values of the coupling and we often know their exact masses. Therefore
a BPS state found at weak coupling in one theory may be continued into
strong coupling and then compared with the BPS states in the other
theory, again at weak coupling. Many states have been matched in this
way between the type IIA on $K3$ and heterotic on $T^4$ \refs{\VafaZH,
\BershadskyQY,\VafaBM}.

The masses of the BPS states vary with the moduli of the theory. In
particular, at certain points in the moduli space they may become
massless. For example, it is well known that certain perturbative
heterotic string states become massless at the points of the enhanced
gauge symmetry in the moduli space. Away from these points their masses
are interpreted simply as arising from the ordinary Higgs mechanism.
On the type IIA side exactly the same points correspond to singular
K3 manifolds \WittenEX, where the singularity arises from
some holomorphic curves inside K3 shrinking to zero size. However,
the particles that become massless are completely non-perturbative
now: they are $D$-branes wrapped around the collapsing cycles
\StromingerCZ. 

From the perspective of the low-energy effective action, the points of
enhanced gauge symmetry are not regular points in the moduli space. In
order to describe the theory in their vicinity we have to add to the action
new fields that are responsible for the appearance of the light
particles. On the heterotic string side this is not difficult since the new
fields are present in the string perturbative expansion. But on the
type IIA side we cannot do the same because the light particles are not
perturbative string states. Instead, we will use the duality to deduce the
type IIA action from the heterotic side.

The extra fields that we have to add to the action are the non-abelian
gauge bosons and scalars. In appendix A we have obtained the
two-derivative part of the effective action for the heterotic string
on $T^4$ that includes these non-abelian fields. In order to apply the
duality transformations we need to know their action on the extra
fields. But, as we mentioned, in the abelian case the gauge fields and
scalars do not transform under duality; we therefore assume the same to
be true in the non-abelian case as well. In appendix B we perform the
duality transformations and obtain the two-derivative part of the
effective action for the type IIA string theory compactified on K3
near the point of enhanced gauge symmetry.

Let us comment on the result, given in \SIIAfinal. First, we already
know that it is fixed by supersymmetry and the field content of the
theory. But by employing S-duality we have written the effective
action in the variables that are simple to interpret for the type IIA
compactification. It also makes it easy for us to transform known
heterotic monopole solutions to the type IIA side. In addition, we
were able to consider only the bosonic sector and avoid the
complications of checking supersymmetry variations.

Second, having found the two-derivative part of the effective action,
we need to find the domain of its validity, and for that we need to
understand the higher-order corrections to the action. From
dimensional analysis the derivative expansion is suppressed by powers
of $\alpha'$, as always. However, what is playing the role of the
coupling constant now?  Ordinarily, in the low-energy effective action
obtained from string theory all interaction vertices contain a power
of the string coupling constant. But it can be easily seen from
\SIIAfinal\ that extra vertices coming from the non-abelian terms do
not contain the dilaton. This is in agreement with the observation in
\StromingerCZ\ that quantum loops of non-perturbative states should
not be suppressed by the string coupling. Hence the non-abelian part
of our action does not have an expansion in the dimensionless coupling
constant and would remain interacting even if the dilaton is taken to
zero.

Using the rules from appendix B, we are now ready to write down a
supergravity monopole which satisfies the equations of motion of the six
dimensional type IIA action at the two derivative level. Type IIA fields
will be primed in the remainder.

Applying S-duality to the fields in the string frame, we get the
following six dimensional type IIA solution:
\eqn\premetric{ \eqalign{
	ds^2 &= \lambda e^{-\Phi}dx^2_\parallel + \lambda
                            e^{\Phi}dx^2_{\perp} \cr
    e^{\phi'} &= e^{- \phi} = \sqrt{\lambda} e^{- \Phi/2}  }}
The only non-zero gauge field is $A^{(3)}$, which is the same as on
the heterotic side
\eqn\aa{ A'^J_i(\vec{r}) = A^J_i(\vec{r}) = C {\scr A}^J_i(C\vec{r}).}
For the anti-symmetric tensor field we simply find $ H'_{ijk} = 0$. Of
the various moduli, $B'_{\alpha\beta}, a'_7, a'_8$ and $a'_9$ are all
constants, whereas
\eqn\aa{ a'_6(\vec{r}) = a_6(\vec{r}) = C\, h(C\vec{r}), \quad
  G'_{\alpha\beta} = G_{\alpha \beta} = {\rm diag}( e^{2\Phi},1,1,1 )  .}
Again $a'_6$ should be interpreted as the Higgs field.  Furthermore,
an expression for $e^{\Phi}$ in terms of the Higgs field was given in
\monopoledilaton.

In order to put the solution in a more recognizable form let us redefine
coordinates as
\eqn\aa{ x_\parallel \to {1\over\sqrt{\lambda}}e^{\Phi_0/2} x_\parallel, \quad
        x_\perp \to {1\over\sqrt{\lambda}} e^{-\Phi_0/2} x_\perp .}
The metric and dilaton are then brought to
\eqn\hetmonopole{\eqalign{ds^2 &= Z^{-1/2} dx^2_\parallel +
                            Z^{1/2}dx^2_{\perp} \cr
        e^{2\phi'} &= e^{2\phi'_0}Z^{-1/2} }}
with the warp factor
\eqn\aa{Z = 1 + {\alpha' C^2 e^{-2\Phi_0}\over 4}
                        (1 -  h^2({1\over\sqrt{\lambda}} e^{-\Phi_0/2}
                       C \vec{r})) }
which goes to one as $r$ goes to infinity.
We would like to express the solution entirely in terms of type II
parameters. The most convenient way of getting an equation for
$C$ is to notice that it is related
to the mass of the $W$-boson on the heterotic side by $M_W = e^{-\Phi_0}
C$, as in \hetWmass. Now apply S-duality to get the type II mass:
\eqn\IIWmass{ M'_W = e^{-\phi'_0} M_W =  e^{-\phi'_0 -\Phi_0} C
    = {1\over\sqrt{\lambda}} e^{-\Phi_0/2} C.}
We will use this equation to eliminate $C$.
Substituting for $C$ and $\Phi_0$ yields the final form of the warp factor:
\eqn\hetwarpfactor{ Z =
 1 + {\alpha' M'^2_W e^{2\phi'_0}\over 4}(1 - h^2(M'_W \vec{r}))
.}
The gauge and Higgs fields can also be obtained from the fiducial solution by:
\eqn\AaFinal{ A^{\prime I}_i(\vec{r}) = {M'_W}\, 
             {\scr A}^I_i(M'_W\vec{r}), \quad
    a^{\prime I}_6(\vec{r}) = {M'_W}\,h^I(M'_W \vec{r})  }
where $a_6^2(\infty) = M'^2_W$. Notice that we have rescaled $a_6$ by
a factor of $e^{-\Phi_0/2}/\sqrt{\lambda}$ here, so in order to use
these formulae with the Lagrangian in appendix B one should also
rescale any field with $\alpha$ or $\beta = 6$ correspondingly. In
particular, $G'_{\alpha \beta}$ becomes
\eqn\Gfinal{
G'_{\alpha \beta} =  {\rm diag}(e^{-2 \phi'_0} Z, 1, 1, 1) \ . }

To summarise, in analogy with the Harvey-Liu monopole, we can start with a
flat space BPS monopole ${\scr A}^J_i(\vec{r}), h^J(\vec{r})$ and use
equations \hetmonopole\ and \hetwarpfactor\ to dress it up with a dilaton
and metric and get a supergravity monopole. The soliton we have written
down lives in low energy supergravity obtained from compactifying type II
string theory on a K3 surface, preserves half of the sixteen supercharges
and makes essential use of the non-abelian gauge fields which in this case
arise from massless D-branes at a special corner of the moduli space of the
compactification.

\subsec{Comparison with the enhan\c con}

As discussed in the introduction, Johnson, Peet and Polchinski \JPP\
wrote down a supergravity solution for the D6-$\overline{\rm D2}$
system using the harmonic function rule.  When reduced to six
dimensions, their solution reads
\eqn\jppmetric{  \eqalign{ds^2_6 &= Z_2^{-{1/2}}Z_6^{-{1/ 2}} dx_\parallel^2 +
          Z_2^{1/2}Z_6^{{1/ 2}} dx_\perp^2 \cr
         e^{2\phi'} &= e^{2\phi'_0}  Z_2^{-1/2} Z_6^{-1/2} }
}
with the following relations:
\eqn\jppdefinitions{
\eqalign{ Z_2 &= 1 - {(2\pi)^4 g N \alpha'^{5/2} \over 2 r V} \cr
             Z_6 &= 1 + { g N \alpha'^{1/2} \over 2 r} \cr
e^{2\phi'_0}  &= {(2\pi\sqrt{\alpha'})^4 g^2\over V} \ .}	}
Here $V$ is the asymptotic volume of the K3, $g$ is the ten dimensional
string coupling, and $N$ is the number of D6-$\overline{\rm D2}$'s.

What is the relation to the metric we have found above? Let us
first consider the case of a charge one monopole. Then from
\singlemonopoledilaton\ we get the warp factor
\eqn\onemonwarp{ Z = 1 + {\alpha' e^{2\phi'_0} \over 4 r^2}
                     (1 - K^2 + 2 H)(M'_W r) .}
An expression for the six dimensional dilaton in terms of ten
dimensional quantities was given in \jppmetric. The $W$-boson in this
case can be identified with a D4-brane wrapped on the K3 surface. As
explained in the introduction, it carries an induced unit of
anti-D0-brane charge and its mass is reduced by the mass of the D0
brane. So we can find the mass immediately:
\eqn\DSixDTwoMW{ M'_W = {2\pi\over 2\pi \sqrt{\alpha'} g}
        \left( {V\over (2\pi\sqrt{\alpha'})^4} - 1 \right) . }
Finally, we would like to drop the exponential terms in $(1 - K^2 + 2
H)(M'_W r)$, which then becomes $2 M'_W r - 1$.  Substituting these in
\onemonwarp\ we find
\eqn\aa{ \eqalign{
    Z &= 1 + {\alpha' e^{2\phi'_0}\over 4 r^2}(2 M'_W r - 1) \cr
 &=
1 - { \alpha'  \tpap^4 g^2
       \over 4 V   r^2 } +
   { g \sqrt{\alpha'} \over 2  r}
      \bigg(1 - {\tpap^4 \over V} \bigg) \cr
  &= \bigg(1 +  { g \sqrt{\alpha'} \over 2 r} \bigg)
     \bigg(1 -  { g \sqrt{\alpha'} \over 2 r}
               {\tpap^4 \over V} \bigg) \cr
    &= Z_2 Z_6 \vert_{N=1} \ . }}
So we conclude that our solution is identical to the one found in
\JPP\ up to exponential corrections of the form $e^{- M'_W r}$,
which smooth out the singularity in the core.

To see what happens for the general solution with arbitrary charge let us
expand $g_{tt} = Z^{-1/2}$ near infinity in powers of $1/r$ where $r$
is the distance from the centre of mass of the monopoles. Then the
$1/r$ and $1/r^2$ terms in the expansion of $g_{tt}$ are proportional
to the tension of the soliton and the square of its charge
respectively. This implies that in the case of $N$ solitons the
leading terms in $Z$ and $Z_2 Z_6$ must agree if they agree for $N =
1$.

Next we would like to briefly discuss the shape of large $N$ monopole
solutions. It is impossible to construct a spherically symmetric, finite
energy Ansatz unless $N = 1$ \WeinbergEQ, but an algorithm for the
construction of axially symmetric monopoles has been proposed
\Prasad. Let us call the axis of symmetry the $x_3$-axis. The
locations of zeros of the Higgs field are usually thought of as the
positions of the monopoles. The solutions of \Prasad\ have a Higgs
field that vanishes only at the origin, so the monopoles are
coincident. Actually, the Higgs field has a zero of order one along
the $x_3$-axis and a zero of order $N$ along the $x_3 = 0$ plane. As a
result one expects the flat region in the interior of the enhan\c con
for $N > 1$ to be of a ``pancake'' type of shape, rather than a
sphere.  We have computed $h^2$ along the $x_3$-axis and on the $x_3 =
0$ plane for these axially symmetric monopoles using expressions from
\PR\ for several N, and the results appear to confirm the above
picture. Since other monopole solutions are lacking, we do not know
how this picture is affected if some of the remaining moduli are used
to deform away from axial symmetry and whether there are solutions
with energy approximately localised on a thin spherical shell.

\newsec{Gauge theory}

\subsec{D-brane picture}

Instead of the D6-$\overline{\rm D2}$ system studied in \JPP, we will
consider a T-dual situation by wrapping $N$ D4-branes on a two-sphere
inside the K3. On a flat D4-brane the low-energy theory is a pure SYM
theory with 16 supersymmetries. When the brane is wrapped on a sphere
inside the K3, the theory becomes twisted \BershadskyQY\ thereby
allowing preservation of 8 supersymmetries. The gauge fields on the
sphere and the complex scalar describing the normal direction to it
inside the K3 have no zero modes, so the massless fields are just a 2+1
dimensional vector multiplet with no hypermultiplets. For $N$ branes
we obtain vector multiplets forming the adjoint of $U(N)$.

As usual, we would like to take a limit in the parameter space such
that only the 2+1 dimensional SYM survives on the brane. In addition
we would like to decouple it from the theory in the bulk. For this we
first have to take $\alpha'$ to zero to get rid of the massive open
string states. Then in order to avoid Kaluza-Klein modes of the
massless fields on the sphere we have to take its area, $A$, to zero.
For the gauge coupling of the resulting 2+1 dimensional theory we have
\Pol
\eqn\gYMthreeD{
{1 \over g^2_{{\rm YM},3}} = {A \over (2 \pi)^2 g\sqrt{\alpha'} }\ , }
which we want to keep constant. Then in order to prevent the string
coupling from blowing up we need to take $A$ to zero as $\sqrt{\alpha'}$ or
faster, but not faster than $\alpha'$, or higher-derivative corrections in
the background fields along the sphere may spoil the gauge theory.  We also
want to avoid higher-derivative corrections in the directions transverse to
the sphere. It means that in these directions the K3 should be flat on a
scale of order $\sqrt{\alpha'}$. Therefore its volume should be bounded
below as
\eqn\VBound{ V \ge \alpha' A \ .}

Now let us discuss decoupling of the brane and the bulk theories.  It
implies that there can be no energy transfer between the two. In fact
it is enough to require that the excitations on the brane cannot
``leak'' out into the bulk. Then by CPT invariance the bulk
excitations cannot influence the brane either. Now, in the limit
$\alpha' \to 0$ the mass of most of the states in the bulk theory goes
to infinity. But the finite energy brane excitations that we are
interested in cannot leak into infinitely massive states. Such bulk
states are therefore always decoupled from the brane. These include
massive closed string states as well as non-perturbative D-brane
states, though we have to be careful with the latter.  While their
tension does go to infinity, some of the states obtained by wrapping
D-branes on very small cycles might remain light. An obvious suspect
is the very same cycle we are putting the D4-brane on.  By wrapping a
D2-brane on it we obtain a W-boson. Fortunately, its mass is given by
\eqn\Wmass{ M'_W = {2 \pi A \over g \tpap^3 } = {1 \over \alpha'
g^2_{{\rm YM},3} } \ , }
which is still infinite in the limit. Volumes of the other cycles in
K3 are determined by the other moduli of the theory. They do not enter
in the field theory on the D4-brane or the monopole solution in the
bulk. We therefore can make them arbitrary and the corresponding
D-brane states decouple from the brane. What we still need is the
decoupling of the massless states as well as the states with masses
remaining finite in the limit $\alpha' \to 0$.  The former are
described by the supergravity Lagrangian in six dimensions and the
latter can only be the higher Kaluza-Klein modes of the
ten-dimensional supergravity on K3.

Let us first analyze the interactions of the massless six-dimensional
supergravity fields with the brane.  For this we look at the string
scattering amplitudes. In our limit the open string diagrams with no
closed string (bulk) vertex operator insertions will be more and more
dominated by the diagrams coming from the 2+1 dimensional gauge
theory, albeit regulated by stringy effects. This is just a statement
that the open string theory on the brane reduces to the gauge
theory. In particular, since $g_{{\rm YM},3}$ is kept fixed in the
limit, we expect the scattering amplitudes to be finite.  The
interaction with the bulk is described by inserting bulk vertex
operators into the open string diagrams. However, every such an
insertion brings with it the corresponding coupling constant. This
coupling constant comes from the six dimensional supergravity that
describes the massless bulk states. For the gravity multiplet it is
the six-dimensional Newton's constant:
$$G_N \sim { g^2 {\alpha'}^4 \over V}\ , $$
where $V$ is the volume of K3. Clearly, it vanishes in our limit. From
\SIIAfinal\ the coupling constant for the gauge fields is just 
$(2 \pi)^3 \alpha'$ and also goes to zero. Now, if we take an open
string diagram that is finite and add a number of closed string
vertices with the coupling constants tending to zero we would find
that the scattering tends to zero, too.

For the higher Kaluza-Klein modes of the ten-dimensional supergravity
on K3 we cannot make the same argument since we do not really know
their couplings. However, when the volume of K3 is small these modes
are heavy and therefore automatically decoupled. When the volume of
the K3 is large we can look at the problem from ten dimensions. The
ten-dimensional Newton's constant and Ramond-Ramond coupling go to
zero in our limit. We will assume that this is enough to decouple the
Kaluza-Klein modes from the D4-brane at arbitrary K3 volume.

\subsec{Soliton picture}

We now want to discuss applying the limits considered in the previous
subsection to the supergravity solution from section two. What we
expect to find are again two decoupled systems: the free supergravity
in the bulk and the theory describing exci\-tations around the core of
the soliton. In the spirit of the AdS/CFT correspondence
\AharonyTI\ we hope to identify the latter with the
gauge theory.

To begin we rewrite the solution in terms of the gauge coupling and
other simple physical parameters. For the gravity multiplet fields we
obtain:
\eqn\IIAmonopole{\eqalign{
  ds^2 &= Z^{-1/2} \, dx^2_\parallel + Z^{1/2} dx^2_\perp \cr
  e^{2\phi'} &= \alpha'  g^4_{{\rm YM},3} {\; A ^2 \over V} Z^{-1/2} 
  \ , \qquad H=0 \cr
  Z &= 1 + {A^2 \over 4 V}\big(1 -  h^2\big(\, \vec{r}\, /
             \alpha' g^2_{{\rm YM},3} \big) \big) \ .
}}
Among the scalars and $U(1)$ gauge fields in our solution that are not
charged with respect to the $SU(2)$ gauge group the only non-trivial
field is $G'_{66} =e^{-2 \phi'_0} Z$. Finally, for the $SU(2)$ gauge
multiplet fields we have:
\eqn\IIAgaugemon{\eqalign{
   A^I_i(\vec{r}) &= {1 \over \alpha' g^2_{{\rm YM},3}}\, 
                   {\scr A}^I_i\big(\, \vec{r}\, /
             \alpha' g^2_{{\rm YM},3} \big) \cr
   a_6^I(\vec{r}) &= {1 \over \alpha' g^2_{{\rm YM},3}}\,
                       h^I \big(\vec{r}\, /
             \alpha' g^2_{{\rm YM},3} \big) \cr
   a_7^I &=  a_8^I =  a_9^I =0 \ .
}}

The first question we should ask ourselves is whether we can trust the
above solution in the limit $\alpha' \to 0$.  Unless $A^2/V$ goes to
infinity as $\alpha'^{-2}$ or faster, this is clearly not the
case. The reason is that the higher-derivative corrections to the
supergravity Lagrangian spoil the solution. Near the core of the
soliton each transverse derivative of the fields above brings with it
a factor of $(\alpha' g^2_{{\rm YM},3})^{-1}$ while it is suppressed
by $\sqrt{\alpha'}$ from the derivative expansion and by a factor from
the metric. However, this factor will not help unless $A^2/V \ge
\alpha'^{-2}$. Therefore the higher the number of derivatives a
term in the supergravity Lagrangian has the bigger its potential
contribution to the action becomes as we take $\alpha' \to 0$.

Taking $A^2/V$ to infinity as $\alpha'^{-2}$ or faster corresponds
either to making the volume of the K3 too small, in violation of
\VBound, or to taking $A$ to infinity.  In both cases we do not have
a clear description of the theory on the D-brane side since the brane
is either embedded in a space highly curved on the scale of
$\sqrt{\alpha'}$ or the Kaluza-Klein modes on the sphere do not
decouple.

The conclusion is that despite having improved on the solution of
\JPP\ we have not been able to find a weakly-coupled dual to the
three-dimensional ${\cal N} =4$ pure gauge theory. This is independent
of whether we take the number of branes, or colors in the gauge
theory, to be large, as the derivative expansion diverges anyway.
However, we can still try to elaborate on what is it we find dual to
the gauge theory.

From the above solution we see that the behavior of the fields in the
$SU(2)$ gauge multiplet depends only on $g^2_{{\rm YM},3}$ and
$\alpha'$, which is consistent with the gauge theory duality. However,
the behavior of all the fields neutral under $SU(2)$ depends on the
ratio $A^2/V$ that can be adjusted, subject to \VBound, independently
of the gauge theory coupling, which suggests that neutral fields have
little to do with the gauge theory.

To investigate, let us consider a limit where we first take the volume
$V$ of the K3 to infinity while keeping the area $A$ of the sphere
constant and only then take $\alpha'$ to zero. From the D-brane side
this is a good limit. The difference is that the bulk theory is now
10-dimensional type IIA string theory on an infinite-volume K3
manifold. Clearly, only the piece of the K3 containing the sphere on which
the D-brane is wrapped is of interest to us. We can therefore think of
the resolved $A_1$ ALE space instead of the K3.  Since the 10-dimensional
couplings vanish as $\alpha' \to 0$, we expect the bulk theory to
decouple from the brane as before.

Taking the volume to infinity in our 6-dimensional supergravity solution
at first does not seem to make much sense. To describe the
10-dimensional geometry via the 6-dimensional Lagrangian we have to
include in it an infinite number of harmonics in the expansion of the
10-dimensional fields over the transverse space. However, the
Lagrangian that we have only contains the lowest harmonics.  Its only
benefit is that it includes inherently 6-dimensional fields which
represent non-perturbative D-brane states responsible for the gauge
symmetry enhancement to $SU(2)$.

The harmonics on the transverse space can be heuristically divided
into two classes: those that have support in the vicinity of the
sphere inside the resolved ALE and those that spread over the entire
space. For instance, the perturbative $U(1)$ part of the $SU(2)$ gauge
multiplet in our Lagrangian contains scalars describing the metric on
the sphere inside the ALE and a gauge field arising from the 3-form in
10 dimensions reduced using the 2-form dual to the sphere. These
harmonics naturally have support in the vicinity of the sphere.  The
non-perturbative part of the $SU(2)$ gauge multiplet can also be
thought of as having support near the sphere. On the other hand, the
6-dimensional gravity multiplet represents 10-dimensional fields
spread over the entire ALE.

Later we intend to take $\alpha' \to 0$. According to our limit, we
will simultaneously take the string coupling $g$ and the area of the
sphere $A$ to zero as well. In a flat 10-dimensional space taking
$\alpha'$ and $g$ to zero would cause the string theory to become
free. In our case the space far away from the sphere is flat and the
harmonics supported over the entire ALE are just plane waves
there. Although we do not know their behavior near the sphere, its
vanishing area suggests that the ``overlap'' between them and the
harmonics supported in the vicinity of the sphere vanishes. We then
expect that the theory in the bulk of the ALE becomes free and
decouples from the 6-dimensonal theory supported in the vicinity of
the sphere. The scale below which the bulk theory is free and
decoupled is set by the 10-dimensional Newton's constant.

Now, the advantage of taking the volume to infinity is that $Z$ goes to 1
and all the fields in our solution that are uncharged under the
$SU(2)$ gauge group become constant. All the excited fields are in the
$SU(2)$ gauge multiplet. The soliton therefore belongs entirely in the
theory supported in the vicinity of the sphere.  The fact that the
metric is not excited means that the soliton is below the decoupling
scale. Indeed, the 10-dimensional Newton's constant is $\sim g^2
\alpha'^4$ while the tension of the soliton is $\sim (\alpha' g_{{\rm
YM},3})^{-2}$. Hence we find that the theory in the bulk of the ALE is not
relevant to the description of the dual gauge theory.

Altogether, the three-dimensional ${\cal N}=4$ $SU(N)$ pure gauge
theory is dual to an $N$- monopole configuration in the
six-dimensional theory living in the vicinity of the sphere in the
resolved $A_1$ ALE space. We know that the low-energy limit of the
latter theory is six dimensional ${\cal N} =2$ $SU(2)$ gauge
theory. The parameters of the theories are related as follows.  The
six-dimensional gauge coupling is just $(2 \pi)^3 \alpha'$, from
\SIIAfinal, and the three-dimensional gauge coupling is related to the
W-boson mass in the six-dimensional vacuum as in \Wmass.  The moduli
space of the three-dimensional theory is the moduli space of $N$
monopoles in six dimensions, with four centre of mass degrees of
freedom removed.

Let us emphasize that we do not know the true monopole configuration.
The solution in \IIAgaugemon\ suffers from $\alpha'$ corrections
whenever $\sqrt{ \alpha'} g^2_{{\rm YM},3}$ is small. However, we know
that the monopole configuration must exist because it carries a
topological charge.  Also note that \IIAgaugemon\ is a good solution
for large $\sqrt{ \alpha'} g^2_{{\rm YM},3}$ and being BPS it must
survive taking $\alpha'$ to zero.

Now, it is well known \OoguriWJ\ that the theory at the singularity of
the $A_{n-1}$ ALE space in type IIA string theory is T-dual to the
little string theory \SeibergZK\ arising on the worldvolume of $n$
NS5-branes in type IIB string theory. Note that at low energies this
little string theory is also described by the ${\cal N} =2$ $SU(n)$
gauge theory.  Its gauge coupling, by duality with the D5-brane, is
again $(2 \pi)^3 \alpha'$.

The picture advocated above is then supported by duality with the
construction of the three-dimensional ${\cal N}=4$ pure gauge theory
due to Hanany and Witten \HananyIE. In that construction the $SU(N)$
gauge theory is realized in type IIB string theory on $N$ D3-branes
stretched between two NS5-branes. If the distance between the
fivebranes is $l$, the gauge coupling in three dimensions will be
$g_{{\rm YM},3}^{-2} = l / (2 \pi)^2 g$, while the mass of the W-boson
in six dimensions, which is just a D1-brane stretched between two
NS5-branes, is $M_W = l / \tpap^2 g$, giving the relation between the
two precisely as in \Wmass.  In the ``supergravity'' picture each
D3-brane stretched between the NS5-branes looks like a magnetic
monopole.  Taking the string coupling to zero while keeping $g_{{\rm
YM},3}$ constant we achieve decoupling of the little string theory on
the NS5-branes from the type IIB string theory in the bulk
\SeibergZK. This yields exactly the same description of the
three-dimensional gauge theory as we have obtained.

\newsec{Discussion}

In this paper we have argued that one can find the correct
non-singular supergravity description of the enhan\c con by solving
supergravity equations with extra light fields added. The resulting
solution is a smooth supergravity monopole. We have also commented on
the fact that known solutions do not correspond to a spherical shell,
even for large $N$.

Subsequently we took a decoupling limit in order to isolate the
Yang-Mills theory living on the D-brane. Unfortunately we have found
that this limit is out of reach for supergravity. Specifically we
found that the derivative expansion diverges. This situation appears
to be generic; in attempts to describe pure Yang-Mills theories with
four supercharges in 3+1 dimensions by means of supergravity it was
also found that $\alpha'$ corrections were large in the decoupling
limit \refs{\KlebanovHB, \MaldacenaYY}. For instance in \MaldacenaYY\
it was found that in order to decouple the QCD scale from the scale of
unwanted Kaluza-Klein modes, one needed to go beyond the supergravity
approximation.

Despite these difficulties, we have been able to distill the duality
in question. We found that the gauge theory is dual to a monopole
configuration in (1,1) little string theory, which is corroborated by
the Hanany-Witten construction.  But it is notable that, unlike
previous cases considered in the literature, gravity turned out to be
completely irrelevant to this duality. The only relevant modes are the
non-gravitational modes that live close to the two sphere inside the
ALE and make up little string theory with a defect.

It is tempting to formulate the result as open/closed duality for little
strings. In the limit $A \to 0,\ g \to 0$ with $A/g$ fixed and $\alpha'$
fixed, one reduces to (1,1) little string theory with an object of finite
tension. Then there would be two pictures of this object, inherited from
the full string theory: in the ``open little string'' picture it is a
``D2'' brane while in the ``closed little string'' picture it can be
described as a Yang-Mills monopole. In the low energy limit $\alpha' \to
0$, the theory living on this ``D2'' brane is three dimensional Yang-Mills
theory with eight supercharges. Unfortunately, keeping the three
dimensional Yang-Mills coupling fixed in this limit causes the divergence
of $\alpha'$ corrections in the closed string picture.

It would be interesting to find out whether the holographic description of
little string theories sheds more light on this situation. Since far away
from the monopoles the little string theory is Higgsed, the double scaling
limit of \GiveonPX\ may be applicable. There the coupling constant for the
dual string description is given by
$${1 \over x} \sim {1 \over M'_W \sqrt{\alpha'}} =  
      \sqrt{\alpha'} g_{{\rm YM},3}^2 \ .$$
It becomes small in the gauge theory limit we are interested in.

In discussing the gauge theory we have not used quite the same system
as in \JPP. Instead of the D6-$\overline{\rm D2}$ system of \JPP\ we
have decided to study the D4 brane wrapped on the sphere inside
K3. The two configurations are related by the $SO(20,4; Z)$ T-duality
group. And while T-duality is supposed to be an exact equivalence of
string theories, naively we do not obtain the same decriptions of the
two cases. Namely, if we look at the relation between the mass of the
W-boson and the 3-dimensional gauge coupling for the D6-$\overline{\rm
D2}$ system we do not obtain the same result \Wmass\ as for the
wrapped D4 brane. The reason is that in the former case the mass of
the W-boson \DSixDTwoMW, which is a D4 brane wrapped on the entire K3,
receives a correction due to the known $R^2$ term in the D-brane
action \BachasUM. However, the 3-dimensional gauge
coupling for the D6 brane wrapped on K3 is naively just
$$ {1 \over g_{{\rm YM},3}^2} = { V \over 2 \pi g \tpap^3} 
   = \alpha' M'_W \bigg(1 - {\tpap^4 \over V} \bigg)^{\! \! -1} \ .$$
The discrepancy may be interpreted as the T-duality prediction for an
additional $R^2 F^2$ term in the D-brane action. When integrated over
K3 such a term would produce a shift in the 3-dimensional Yang-Mills
coupling that would restore consistency with T-duality.  One can in
fact explicitly show the presence of this term and other similar ones
\WZ.

\bigskip
\bigskip

\centerline{\bf  Acknowledgements:}

It is a pleasure to thank N. Arkani-Hamed, M. Gutperle, J. Maldacena,
S. Minwalla,  P. van Nieuwenhuizen, B. Pioline and M. Ro\v{c}ek for
discussions.

The research of MPW was supported in part by NSF grant DMS-9709694. The
research of SZ is supported in part by NSF grant PHY-0098527.

\bigskip

\appendix{A} {Non-abelian toroidal reduction}

In this appendix we present formulae for the dimensional reduction of
the theory with the Lagrangian of the form \SHetFull\ with non-abelian
gauge fields.  We will closely follow the work of Maharana and Schwarz
\MaharanaMY.

We consider a theory in $D+d$ dimensions compactified on a
$d$-dimensional torus. All fields will be independent of the
coordinates $y^\alpha$ along the torus. To avoid confusing notation in
this appendix only we will add a hat to all $(D+d)$-dimensional
quantities while their cousins in $d$ dimensions remain intact. Then
the $(D+d)$-dimensional index $\hat \mu$ splits into a $D$-dimensional
$\mu$ and an internal $d$-dimensional $\alpha$.

For the purposes of dimenional reduction let us explicitly rewrite
\SHetFull\ in the new notation
\eqn\Cfullaction{\eqalign{
S = {2 \pi \over \tpap^8} & \int dx ~ \int_{T^d} dy~ \sqrt{- \hat g}~ 
e^{-2 \hat\Phi}\Big\{ \hat  R(\hat g) +  4 \hat g^{\hat \mu \hat \nu} 
\partial_{\hat \mu} \hat\Phi \partial_{\hat \nu} \hat\Phi  \cr &  
- {1 \over 12}\hat g^{\hat \mu \hat \mu '}~  \hat g^{\hat \nu \hat \nu '}
~ \hat g^{\hat \rho \hat \rho '} ~ \hat H_{\hat \mu \hat \nu \hat \rho} ~
\hat H_{\hat \mu' \hat \nu' \hat \rho'} 
- {\alpha' \over 8} \hat g^{\hat \mu \hat \rho} \hat g^{\hat \nu \hat \lambda}
\tr \big( \hat F_{\hat \mu \hat \nu} ~ 
            \hat F_{\hat \rho \hat \lambda} \big) \Big\}. }}
For completeness we also write down our conventions. The non-abelian
gauge field strength is
\eqn\CFdefn{ \hat F_{\hat \mu \hat \nu} =
 \partial_{\hat \mu} \hat A_{\hat
\nu} - \partial_{\hat \nu} \hat A_{\hat \mu} +
\big[\hat A_{\hat\mu} \, , \hat A_{\hat\nu}\big] ~, }
and $\hat H_{\hat \mu \hat \nu \hat \rho}$ includes Chern-Simons form
 as follows:
\eqn\CHdefn{\hat H_{\hat \mu \hat \nu \hat \rho} = 
  \partial_{\hat \mu} \hat B_{\hat \nu \hat \rho} - 
{\alpha' \over 4} \tr \big( \hat A_{\hat\mu} 
 \hat F_{\hat\nu \hat\rho} - {1\over 3}
   \hat A_{\hat\mu} \big[ \hat A_{\hat\nu} \, , 
      \hat A_{\hat\rho} \big] \big)
 + ({\rm cyc.~ perms.})
} 

Now we give definitions of the $d$-dimensional fields. The
($D+d$)-dimensional metric $\hat g_{\hat \mu \hat \nu}$ gives rise to
the metric $g_{\mu \nu}$ in $D$ dimensions, $d$ of $U(1)$ gauge fields
$A^{(1)\alpha}_\mu$ and a $d \times d$ symmetric matrix of scalars
$G_{\alpha \beta}$. We define these fields as in \MaharanaMY; first we
use the local Lorentz invariance to bring the $(D+d)$-dimensional
vielbein into a triangular form:
\eqn\fullvielbein{\hat e^{\hat r}_{\hat \mu} = \left(\matrix {e^r_{\mu} &
A^{(1)\beta}_{\mu} E^a_{\beta}\cr 0 & E^a_{\alpha}\cr}\right ) \ ,}
where $\hat r = \{r, a\}$ are vielbein indices.  In the above relation
we have already defined the gauge fields.  The internal metric of the
torus, $G_{\alpha \beta} = E^a_{\alpha} \delta_{ab} E^b_{\beta}$,
becomes the symmetric matrix of scalars in $D$ dimensions while the
``spacetime'' metric is $g_{\mu \nu} = e^r_{\mu}\eta_{rs}
e^s_{\nu}$. Then the complete $(D + d)$-dimensional metric can be
written as
\eqn\fullmetric{\hat g_{\hat \mu \hat \nu} = \left (\matrix {g_{\mu \nu} +
A^{(1)\gamma}_{\mu}G_{\gamma\delta} A^{(1)\delta}_{\nu} &
A^{(1)\gamma}_{\mu}G_{\gamma \beta}\cr A^{(1)\gamma}_{\nu}G_{\gamma
\alpha} & G_{\alpha \beta}\cr}\right ) \ .}

Next we define the $D$-dimensional dilaton field $\phi$
\eqn\CDdilaton{e^{-2 \phi} = 
{ V \over \tpap^4} e^{-2 \hat \Phi} = 
{\int_{T^d} dy \, \sqrt{G} \over \tpap^4} e^{-2 \hat \Phi} ,}
where $V$ is the volume of the internal torus.  We have absorbed the
fourth power of $\tpap$ in the above for our later convenience in the
main text.

The $(D+d)$-dimensional gauge fields $\hat A^I_{\hat \mu}$ upon
compactification give rise to the non-abelian gauge fields
$A^{(3)I}_\mu$ together with $d$ adjoint scalars $a^I_\alpha$. Their
definitions are exactly the same as in the abelian case \MaharanaMY
\eqn\CAdefn{\eqalign{
&a^I_\alpha = \hat A^I_\alpha \cr
&A^{(3)I}_\mu = \hat A^I_\mu -  a^I_\alpha \, A^{(1)\alpha}_\mu \ .}}

Finally, from the $\hat B_{\hat \mu \hat \nu}$ field we obtain in $D$
dimensions the $d \times d$ anti\-symmetric matrix of scalars
$B_{\alpha \beta}$, $d$ of $U(1)$ gauge fields $A^{(2)}_{\mu
\alpha}$ and the $D$-dimensional two-form field $B_{\mu \nu}$. Their
definitions differ only slightly from those in \MaharanaMY
\eqn\CDBdefn{\eqalign{
&B_{\alpha \beta} = \hat B_{\alpha \beta} \cr
&A^{(2)}_{\mu\alpha} = \hat B_{\mu \alpha} - 
     B_{\beta \alpha} A^{(1) \beta}_\mu + 
        {\alpha' \over 4} \tr ( A^{(3)}_\mu a_\alpha) \cr
&B_{\mu \nu} = \hat B_{\mu \nu} +{1 \over 2}
\big( A^{(1)\alpha}_\mu A^{(2)}_{\nu \alpha} -  
      A^{(1)\alpha}_\nu A^{(2)}_{\mu \alpha} \big) + {\alpha' \over 4}
\big(\tr(A^{(3)}_\mu a_\gamma )\, A^{(1)\gamma}_\nu -
     \tr(A^{(3)}_\nu a_\gamma )\, A^{(1)\gamma}_\mu \big) \ .
}}

As for the $D$-dimensional field strength $H$, it now includes
Chern-Simons terms for all the gauge fields
\eqn\CDHdefn{\eqalign{
H_{\mu \nu \rho} = \del_\mu B_{\nu \rho} - {\alpha' \over 4} &
\tr\big( A^{(3)}_{\mu} F^{(3)}_{\nu \rho} - {1\over 3} A^{(3)}_{\mu}
\big[A^{(3)}_{\nu}, A^{(3)}_{\rho}\big]\big) \cr -{1 \over 2} &\big(
A^{(1)\alpha}_\mu F^{(2)}_{\nu \rho \alpha} + A^{(2)}_{\mu \alpha}
F^{(1)\alpha}_{\nu \rho} \big) + ({\rm cyc.~ perms.}) \ .  }}

With the above definitions after a somewhat tedious calculation we
find the following action for the dimensionally reduced theory
\eqn\CDaction{\eqalign{
S_D = &{2 \pi \over \tpap^4} \int dx ~ \sqrt{ -g} e^{-2\phi} \Big\{
R(g) +4 g^{\mu \nu} \partial_{\mu} \phi \partial_{\nu} \phi
+ {1 \over 4} g^{\mu \nu} \partial_{\mu} G_{\alpha \beta} 
                          \partial_{\nu} G^{\alpha \beta} \cr
&-{1 \over 4} g^{\mu \rho} g^{\nu \lambda} G_{\alpha
\beta} F^{(1)\alpha}_{\mu \nu} F^{(1) \beta}_{\rho \lambda}
 -{1 \over 12} (H_{\mu \nu \rho})^2
 -{1 \over 4} (H_{\mu \nu \alpha})^2 
 -{1 \over 4} (H_{\mu \alpha \beta})^2 
 -{1 \over 12} (H_{\alpha \beta \gamma})^2 \cr
&-{\alpha' \over 8}
 \tr(F^{(3)}_{\mu \nu} + a_\alpha \, F^{(1)\alpha}_{\mu \nu})^2 
 -{\alpha' \over 4} \tr(D_\mu a^I_\alpha)^2
 -{\alpha' \over 8} \tr([ a_\alpha, a_\beta])^2 \Big\} \ ,
}} 
where $D_\mu = \del_\mu + [A^{(3)}_\mu, \cdot \,]$ denotes the gauge
covariant derivative and the implied raising and lowering of the
indices is done with $(g^{-1})^{\mu \nu}$ and $(G^{-1})^{\alpha
\beta}$. We have also defined the following quantities
\eqn\CDextradefn{\eqalign{
&H_{\mu \nu \alpha} = F^{(2)}_{\mu \nu \alpha} - 
      {\alpha' \over 2} \, \tr( a_\alpha \, F^{(3)}_{\mu \nu}) - 
        C_{\alpha \beta}F^{(1)\beta}_{\mu \nu} \cr
&H_{\mu \alpha \beta} = \del_\mu B_{\alpha \beta} +
 {\alpha' \over 4} \big\{ \tr(a_\alpha D_\mu a_\beta) -
                          \tr(a_\beta  D_\mu a_\alpha) \big\} \cr
&H_{\alpha \beta \gamma} = -{\alpha' \over 2} 
            \tr( a_\alpha [a_\beta, a_\gamma]) \ ,}}
where
\eqn\CCdefn{ C_{\alpha \beta} = B_{\alpha \beta} 
               + {\alpha' \over 4} \tr(a_\alpha a_\beta)  \ . }

\appendix{B}{S-duality transformations}

In this appendix we will explicitly apply S-duality \refs{\SenCJ,
\WittenEX}\ to the action \CDaction\ specialized for the heterotic
string on a four-torus. In our conventions the duality transformations
are particularly simple \Pol. First, the string tension is the same
for the two theories, which can be checked in the end by transferring
to the heterotic side the tension of the string obtained by wrapping
the NS5-brane on K3. Next, the dilaton and the string metric change
as
\eqn\SDphig{ \phi' = -\phi, \qquad g'= e^{-2 \phi} g \ ,}
where we have denoted the type IIA six-dimensional fields with a
prime.

Making this change of variables in the heterotic action we obtain
\eqn\SIIAone{\eqalign{
S_{\rm IIA} = &{2 \pi \over \tpap^4} \int d^{\, 6 \!}x ~ 
\sqrt{-g'} \Big\{ e^{-2 \phi'}\big( R(g') 
+4  \del_\mu \phi' \del^\mu \phi' \big)
+ {1 \over 4}  e^{-2 \phi'}(\del_\mu G_{\alpha \beta})^2 \cr
&-{1 \over 4}(F^{(1)\alpha}_{\mu \nu})^2
 -{1 \over 12}  e^{2 \phi'} (H_{\mu \nu \rho})^2
 -{1 \over 4} (H_{\mu \nu \alpha})^2 
 -{1 \over 4}  e^{-2 \phi'} (H_{\mu \alpha \beta})^2 
 -{1 \over 12}  e^{-4 \phi'}(H_{\alpha \beta \gamma})^2 \cr
&-{\alpha' \over 8}
 \tr(F^{(3)}_{\mu \nu} + a_\alpha \, F^{(1)\alpha}_{\mu \nu})^2 
 -{\alpha' \over 4}  e^{-2 \phi'} \tr(D_\mu a^I_\alpha)^2
 -{\alpha' \over 8}  e^{-4 \phi'} \tr([ a_\alpha, a_\beta])^2 \Big\} \ .
}}
All the quantities here are defined by the same expressions
\CDHdefn, \CDextradefn\ and \CCdefn\ as in the Heterotic compactification. That
is so because none of these definitions depend on the metric.

Now we need to dualize the six-dimensional three-form field strength
$H$. Recall its definition from \CDHdefn
\eqn\Hshort{H = dB - \omega^{cs} \ , }
where $\omega^{cs}$ denotes a full Chern-Simons form from
\CDHdefn. Then we have 
\eqn\dH{dH = -d \omega^{cs} = - {\alpha' \over 4} \tr(F^{(3)} \wedge F^{(3)})
  - F^{(1)\alpha} \wedge F^{(2)}_\alpha \ . }
Note that unlike $\omega^{cs}$ itself, its exterior derivative is
globally well-defined and gauge-invariant.  We can then treat $H$ as
an independent field if we add to the action a Lagrange multiplier
that will enforce \dH\ as a constraint, namely
\eqn\HLagrange{
\delta S_{\rm IIA} = {2 \pi \over \tpap^4} \int B' \wedge( dH + d
\omega^{cs} ) \ ,}
where $B'$ is another 2-form field.

Now we will focus only on those terms in the action that  contain $H$:
\eqn\ScontH{
S_{{\rm IIA}|H} =  - {2 \pi \over \tpap^4} \int \Big( {1 \over 2} 
e^{2 \phi'} \, H \wedge \star' H - B' \wedge ( dH + d\omega^{cs} ) \Big) \ .}
Integrating out $B'$ just gives back the original action
\SIIAone. Instead, integrating out $H$, we need to set it to the
solution of the equations of motion
\eqn\Heqmot{
e^{2 \phi'} \star' H - dB' =0 \ .}
Substituting this back in \ScontH\ we obtain the action in terms of
the dual field $B'$
\eqn\SIIAtwo{S_{{\rm IIA}|H} =  - {2 \pi \over \tpap^4}\int \Big(
{1 \over 2} e^{-2\phi'} dB' \wedge \star' dB' - B' \wedge
 d\omega^{cs}  \Big) \ .}
The above is the correct action for the $B$ field of the type IIA
compactified on K3, including the normalization of the kinetic
term and the periodicity of the $B$ field \VafaFJ.

Altogether, dropping the primes on the six dimensional type IIA
fields, we have the following action
\eqn\SIIAfinal{\eqalign{
S_{\rm IIA} = &{2 \pi \over \tpap^4} \int d^{\, 6 \!}x ~ 
\sqrt{-g} \Big\{ e^{-2 \phi}\Big( R(g) 
+4  \del_\mu \phi \del^\mu \phi 
+ {1 \over 4}  (\del_\mu G_{\alpha \beta})^2
 -{1 \over 12} (H_{\mu \nu \rho})^2 \cr
&-{1 \over 4} (H_{\mu \alpha \beta})^2 
 -{\alpha' \over 4} \tr(D_\mu a^I_\alpha)^2 \Big)    
 -{1 \over 4}(F^{(1)\alpha}_{\mu \nu})^2
 -{1 \over 4} (H_{\mu \nu \alpha})^2 \cr
&-{\alpha' \over 8}  
    \tr(F^{(3)}_{\mu \nu} + a_\alpha \, F^{(1)\alpha}_{\mu \nu})^2 
 -{1 \over 12}  e^{-4 \phi}(H_{\alpha \beta \gamma})^2
 -{\alpha' \over 8}  e^{-4 \phi} \tr([ a_\alpha, a_\beta])^2
   \Big\}\cr
&+{2 \pi \over \tpap^4}\int B \wedge \Big( 
   {\alpha' \over 4} \tr(F^{(3)} \wedge F^{(3)})
    + F^{(1)\alpha} \wedge F^{(2)}_\alpha \Big) \ . }}
On this side the definition of $H$ does not include the Chern-Simons
term anymore
$$H_{\mu \nu \rho} = \del_\mu B_{\nu \rho} + ({\rm cyc.~perms.}) 
   \qquad  {\rm or}\ \, H=dB \ ,$$
while the definitions of $H_{\mu \nu \alpha}$,
$H_{\mu \alpha \beta}$ and $H_{\alpha \beta \gamma}$ are given by the
same expressions as before, \CDextradefn\ and \CCdefn.

\listrefs 
\bye